\documentclass[preprint,review,12pt]{elsarticle}

\usepackage{amssymb}
\journal{Physics Letters B}

\def\barray{\begin{eqnarray}}
\def\earray{\end{eqnarray}}
\def\beq{\begin{equation}}
\def\eeq{\end{equation}}

\begin{document}

\begin{frontmatter}

\title{Entanglement Renormalization and Two Dimensional String Theory}


\author{J. Molina-Vilaplana}
\ead{javi.molina@upct.es}
\address{Universidad Polit\'ecnica de Cartagena. C/Dr Fleming S/N. 30202. Cartagena. Spain}
%This article is registered under preprint number: /hep-th/1510.09020

\begin{abstract}
The entanglement renormalization flow of a (1+1) free boson is formulated as a path integral over some auxiliary scalar fields. The resulting effective theory for these fields amounts to the dilaton term of non-critical string theory in two spacetime dimensions. A connection between the scalar fields in the two theories is provided, allowing to acquire novel insights into how a theory of gravity emerges from the entanglement structure of another one without gravity.

\end{abstract}


\end{frontmatter}

\section{Introduction}
\label{intro}
Currently, striking connections between the spacetime structure in gravitational theories and patterns of entanglement in dual quantum theories have emerged \cite{ryutak,ent_geom1,ent_geom2,firstlaw1,firstlaw2}. These incipient insights have been mostly understood in the framework of the AdS/CFT correspondence \cite{adscftbib2,adscftbib3,adscftbib4}. The holographic formula of the entanglement entropy \cite{ryutak} is a dazzling manifestation of these connections. It has been also noteworthy to observe how hyperbolic geometries come associated to the entanglement renormalization tensor networks (MERA) \cite{vidalmera} used in numerical investigations of the ground states of quantum critical systems \cite{swingle}. Using MERA and particularly its continuous version, cMERA \cite{cMERA}, geometric descriptions of relevant states in field theories have been provided \cite{taka1, prog1, molina015}. However, it has not been possible to establish if these geometrical representations correspond to solutions of any known theory of gravity. 

Our objective in this Letter is to provide a simple example in which the cMERA representation of a free (1+1) dimensional quantum field theory can be described in terms of the solutions of a gravity theory. As usual in physics, useful information can be gained by considering low-dimensional models. Here, we find that the cMERA representation of the ground state of a free massive boson amounts to a known solution of string theory in two spacetime dimensions. This theory, despite being the 'simplest' string theory, retains many interesting features of its more complex peers in higher dimensions and remarkably, it can be nonperturbatively formulated in terms of a model of nonrelativistic fermions via the $c=1$ matrix model \cite{matrix_model}.

\section{Entanglement Renormalization for QFT}
\label{cMERA}
The Multi-Scale Entanglement Renormalization Ansatz (MERA)  \cite{vidalmera, cMERA} is a real-space renormalization group procedure on the quantum state which represents the wavefunction of the quantum system (usually in its ground state) at different length scales labeled by $u$. In MERA, $u=0$ usually refers to the state at short lenghts (UV-state $|\Psi_{UV}\rangle$). In general, this state is highly entangled and acts as a starting reference point for the renormalization flow. MERA carries out a renormalization transformation at each length scale $u$ in which, prior to coarse graining the effective degrees of freedom at that scale, the short range entanglement between them is unitarily removed through a {\em disentangler}. 
%Therefore, MERA iteratively removes the quantum correlations between small adjacent regions of space at each length scale. 
The procedure is applied an arbitray number of times until the IR-state $|\Psi_{IR}\rangle$ is reached \footnote{For massive theories, $|\Psi_{IR}\rangle$ is a completely unentangled state. In massles CFT, $|\Psi_{IR}\rangle$ amounts to the entangled vacuum of the theory.}.  

The MERA flow can be implememnted in a reverse way: starting from $|\Psi_{IR}\rangle$, it works by unitarily adding entanglement at each length scale until the correct $|\Psi_{UV}\rangle$ is generated. To fix the concept, let us generate the state $|\Psi(u)\rangle$ obtained by adding some amount of entanglement between left and right propagating modes of momentum $|k| \leq \Lambda e^{-u}$ to the state $|\Psi_{IR}\rangle$, 
\beq
\label{eq1.1}
|\Psi(u)\rangle = P\, e^{-i\int_{u_{IR}}^{u} d\hat{u}\, (\mathcal{K}(\hat{u}) + \mathcal{D})}\, |\Psi_{IR}\rangle.
\eeq
The symbol $P$ is a path ordering operator which allocates operators with bigger $u$ to the right and $\Lambda$ is a UV momentum cut-off. The operator $\mathcal{K}(\hat{u})$ creates a definite amount of entanglement at a given scale $u$ and, in its most general form can be written as,
\beq
\label{eq1.2}
\mathcal{K}(\hat{u})=\int d^d k\, \Gamma(k/\Lambda)\, g(\hat{u},k)\, \mathcal{O}_k,
\eeq
where $\mathcal{O}_k$ is an operator acting at the energy scale given by $k$ and $\Gamma(x)=1$ for $0<x<1$ and zero otherwise. The function $g(\hat{u},k)$ depends on the state and the model that one deals with and represents the strenght of the entangling process at a given scale. The operator $\mathcal{D}$ corresponds to coarse-graining \cite{cMERA,taka1}. To focus only in the entanglement flow along cMERA while avoiding the effects of the coarse graining process it is useful to rescale the cMERA states as,
\beq
\label{eq1.3}
|\widetilde{\Psi}(u)\rangle = e^{iu\, \mathcal{D}}\,|\Psi(u)\rangle= P\, e^{-i\int_{u_{IR}}^{u} d\hat{u}\, \widetilde{\mathcal{K}}(\hat{u})}\, |\Psi_{IR}\rangle.
\eeq
Now, the \textit{entangler} operator is given in the \textit{interaction} picture,
\beq
\label{eq1.4}
 \widetilde{\mathcal{K}}(\hat{u}) = e^{-i\hat{u}\mathcal{D}}\, \mathcal{K}(\hat{u})\, e^{i\hat{u}\mathcal{D}}= \int d^d k\, \Gamma(k\, e^{\hat{u}} /\Lambda)\, g(\hat{u},k\, e^{\hat{u}})\, \widetilde{\mathcal{O}}_k,
\eeq
with $ \widetilde{\mathcal{O}}_k = e^{-i\hat{u}\mathcal{D}}\, \mathcal{O}_k\, e^{i\hat{u}\mathcal{D}}=e^{-d \hat{u}}\, \mathcal{O}_{k\, e^{\hat{u}}} $. 

This Letter will consider the ground state of a $d=1$ free bosonic theory with an action given by,
\beq
\label{eq1.5}
S=\int dt dx \left[\left(\partial_t\, \phi \right)^2 + \left(\partial_x\, \phi \right)^2 -m^2 \phi^2 \right].
\eeq 
For this model, $\widetilde{\mathcal{K}}$ reads as \cite{taka1},
\beq
\label{eq1.6}
 \widetilde{\mathcal{K}}(\hat{u}) = -\frac{i}{2} \int dk\, \left( g_k(\hat{u})\, a^{\dagger}_k\, a^{\dagger}_{-k} - g_k(\hat{u})^{*}\, a_k\, a_{-k}\right),
\eeq
with $g_k(\hat{u})= \Gamma(k e^{\hat{u}}/\Lambda)\, g(\hat{u},k)$. The operators $a^{\dagger}_k, a_k$ are defined as the creation and anihilation operators of a field mode with momentum $k$ with respect to $|0\rangle$, the ground state of the theory at $u=0$.  The conmutation relations are $\left[ a_k, a^{\dagger}_p \right] = \delta({k- q)}$, and zero otherwise. With this, the cMERA state $|\widetilde{\Psi}(u)\rangle$ amounts to the SU(1,1)/ U(1) generalized coherent state \cite{perelmonov},
\beq
\label{eq1.7}
|\Phi\rangle =  \mathcal{N}\, \exp\, \lbrace -\frac{1}{2}\int dk\, \left[\Phi_k(u)\, K_{+} - \overline{\Phi}_k(u)\, K_{-}\right]\rbrace\, |0 \rangle,
\eeq
with $\Phi_k(u) = \int_{0}^{u}  g_k(\hat{u})\, d\hat{u}$, $\overline{\Phi}_k(u)\equiv \Phi^{*}_k(u)$ and a normalization constant given by $\mathcal{N}=\exp\lbrace-1/2\, \int dk\, |\Phi_k(u)|^2 \rbrace$. The bilinear bosonic operators defined by
\beq
\label{eq1.10}
K_{+} = a^{\dagger}_k\, a^{\dagger}_{-k}, \quad K_{-} = a_k\, a_{-k},
\eeq
together with $K_0 = \frac{1}{2}(a_k^{\dagger}\, a_{k} + a_{-k}^{\dagger}\, a_{-k} + 1)$, satify the Lie algebra conmutation relations of the group SU(1,1)
\beq
\label{eq1.11}
\left[ K_{0},\, K_{\pm}\right] =\pm K_{\pm}\, \quad \left[ K_{-},\, K_{+}\right] =2 K_{0},
\eeq
and
\beq
\label{eq1.11x}
 K_{-}\, |\Phi\rangle = \Phi_k(u)\, |\Phi\rangle, \quad \langle \Phi|\, K_{+} = \overline{\Phi}_k(u)\, \langle \Phi|.
\eeq

From this point of view, the cMERA flow amounts to a sequential generation of a set of coherent states $ |\, \Phi\rangle$  where the state $|0\rangle$ acts as the reference state \footnote{We refer to \cite{molina015} for an analysis of the differential generation of entanglement required to construct the set of cMERA coherent states (\ref{eq1.7})}. This set of coherent states satisfy,
%\beq
%\label{eq1.11y}
%\langle \Phi'|\, \Phi\rangle = \exp\left[-\frac{1}{2} \int dk \left(\, |\, \Phi'|^2 +|\, \Phi|^2 -2\overline{\Phi'}\Phi\, \right) \right],
%\eeq
%and
\beq
\label{eq1.12}
\int\, d\mu(\Phi)\,  |\Phi\rangle \langle \Phi| = \mathbf{I},
\eeq
where $d\mu(\Phi)$ is the SU(1,1)-invariant Haar measure on SU(1,1)/ U(1). Furthermore, each one of these states are one-to-one corresponding to the points in the coset SU(1,1)/U(1) manifold except for some singular points \cite{provost}. Namely, the states $|\Phi\rangle$ are embeded into a topologically nontrivial space corresponding to a 2-dimensional hyperbolic space. In other words, each cMERA state $|\Phi\rangle$  corresponds to a point on a two dimensional hyperbolic space. It may be argued that once provided a suitable measure of the distance between the states $|\Phi\rangle$, then a geometric description of the cMERA renormalization flow should correspond to the metric of a two dimensional AdS space \cite{molina015}. More to be said about this point later in this work in which, we turn to ask whether the cMERA renormalization flow for the model (\ref{eq1.5}) may be considered in terms of a concrete gravitational theory (see also \cite{zyl15}).

\section{cMERA Path Integral and Effective Action}
\label{pathint}
Here, we formulate cMERA as a path integral using the coherent state formalism. To this aim, we consider the amplitude
\beq
\label{eq2.1}
G(u_F,u_{IR})=\langle \Phi_{F}\, |P\, \exp\, \lbrace\,  -i\int_{u_{IR}}^{u_{F}} d\hat{u}\, \widetilde{\mathcal{K}}(\hat{u})\rbrace\,   |\Phi_{IR}\rangle.
\eeq
Recalling that $ \partial_u\, \Phi_k(u)=g_k(u)$, then if one follows the standard procedure of dividing the renormalization scale interval $(u_{F} - u_{IR})$ into $N$ intervals, each with $\epsilon = (u_{F}  -u_{IR})/N$, then inserting the resolution of identity (\ref{eq1.12}) at each interval point \footnote{We also must note that the transition amplitude between two different coherent states (\ref{eq1.7}) is given by $\langle \Phi'|\, \Phi\rangle = \exp\left[-1/2 \int dk \left(\, |\, \Phi_k(u')|^2 +|\, \Phi_k(u)|^2 -2\, \overline{\Phi}_k(u')\, \Phi_k(u)\, \right) \right]$.}, and finally letting $N$ go to infinity while dropping $\mathcal{O}(\epsilon^2)$ terms, the amplitude (\ref{eq2.1}) can be written as a formal generalized coherent state path integral,
\beq
\label{eq2.2}
G(u_F,u_{IR})= \int d\mu(\Phi,\overline{\Phi})\, \exp\, \lbrace i\,  \mathcal{S}_{{\rm eff}}[\Phi,\overline{\Phi}] \rbrace,
\eeq
where
\barray
\label{eq2.3}
 \mathcal{S}_{{\rm eff}}[\Phi,\overline{\Phi}]&=&-\int_{u_{IR}}^{u_F} du\, \left[\mathcal{L}[\Phi,\overline{\Phi}; u]+\langle \Phi\, |\, \widetilde{\mathcal{K}}(u)|\, \Phi\rangle \right], \\ \nonumber
\mathcal{L}[\Phi,\overline{\Phi};u]&=&  \frac{1}{2i} \int dk\, \left[\overline{\Phi}_k(u)\, \partial_u\Phi_k(u)-\Phi_k(u)\, \partial_u\overline{\Phi}_k(u)\right].
\earray
We have explicitly dropped out the projection operators onto the initial and final states but it must be noted that the Euler-Lagrange equations derived from $\mathcal{S}_{{\rm eff}}[\Phi,\overline{\Phi}]$ are accompanied by the boundary conditions $\Phi_k(u_F) \equiv \Phi_k(u_N)$ and $\Phi_k(u_{IR}) \equiv \Phi_k(u_0)$ respectively. Regarding this, the effective action only contains two terms. The second term is tantamount to the matrix element of the entangler operator $\widetilde{\mathcal{K}}$ in the coherent state basis while the first term $\mathcal{L}[\Phi,\overline{\Phi};u]$ is pure geometric;  it is indeed a Berry phase that describes how the quantum entanglement is created along the cMERA flow. Using the expressions (\ref{eq1.6}), (\ref{eq1.10}) and (\ref{eq1.11x}), it can be shown that $\mathcal{L}[\Phi,\overline{\Phi};u] = \langle \Phi\, |\widetilde{\mathcal{K}}(u)|\, \Phi\rangle$, so $ \mathcal{S}_{{\rm eff}}[\Phi,\overline{\Phi}]$ totally accounts for the quantum fluctuations along the cMERA flow and can be written as,
\barray
\label{eq2.4}
 \mathcal{S}_{{\rm eff}}[\Phi,\overline{\Phi}] &=& i\, \int_{u_{IR}}^{u_F} du\, dk\, \left[\, \overline{\Phi}_k(u)\, \partial_u\Phi_k(u)-\Phi_k(u)\, \partial_u\overline{\Phi}_k(u)\, \right]  \\ \nonumber
 &=&-2 \int_{u_{IR}}^{u_F} du\, dk\,\left[  \overline{\Phi}_k(u)\, \partial_u\Phi_k(u)\right] .
\earray
 
 Here, we will be mainly interested in the amplitude $G(0,u_{IR})$, i.e, the amplitude whose effective action $\mathit{S}_{{\rm eff}}[\Phi,\overline{\Phi}]$ relates to the full cMERA renormalization coordinate $u$. Furthermore, we turn to a real-space description instead of the $k$-space one such that,
 \beq
 \label{eq2.5}
 \mathit{S}_{{\rm eff}}[\Phi,\overline{\Phi}] = -2 \int du\, dx\, \left[ \overline{\Phi}(u;x)\, \partial_u\Phi(u;x)\right].
 \eeq
This result has been derived using the \textit{interaction} picture. Nevertheless, to portray the full cMERA renormalization flow one might also take into account the coarse graining process, which amounts to recover the Schr\"odinger picture of cMERA (eq.(\ref{eq1.1})). Noting that  $\widetilde{\mathcal{K}}(u)=e^{-u}\, \mathcal{K}(u)$ (see eq.(\ref{eq1.4})),  this can be used to show that the full cMERA flow can be expressed in terms of a path integral with an effective action given by,
\beq
\label{eq2.6}
\mathit{S}_{{\rm eff}}[\Phi,\overline{\Phi}] = -2 \int du\, dx\, \left[ \overline{\Phi}(u;x)\, e^{-u}\, \partial_u\Phi(u;x)\right] .
\eeq

As stated above, in \cite{taka1,molina015} it has been shown that the coherent state description of  cMERA for the model (\ref{eq1.5}) yields a natural geometric representation of the renormalization flow by means of a two dimensional metric defined on a manifold with coordinates $u$ and $x$.  This is given by,
\beq
\label{eq2.7}
ds^2=g_{uu}\, du^2 + e^{-2u}\, dx^2,
\eeq
with $g_{uu} = g_k(u)^2$. For the ground state of the free scalar theory with mass $m$, the cMERA variational parameter $\Phi_k(u)$ is obtained by  minimizing the energy density $\mathcal{E}=\langle \Psi_{IR}\, |\, \mathcal{H}(u_{IR})|\, \Psi_{IR}\rangle$, with $\mathcal{H}$ the Hamiltonian function derived from (\ref{eq1.5}). The result reads as \cite{cMERA,taka1},
\beq
\label{eq2.8}
\Phi_k(u)=\left[ -\frac{1}{4}\, \log\, \frac{k^{2} + m^{2}}{\Lambda^{2} + m^{2}}\right]_{k=\Lambda e^{-u}} =-\frac{1}{4}\, \log\, \frac{e^{-2u} + \overline{m}^2}{1 + \overline{m}^2},
\eeq
with $\overline{m} = m/\Lambda \ll 1$. This variational solution inmediatly leads to,
\beq
\label{eq2.9}
g_k(u)=\partial_u\, \Phi_k(u) =\frac{1}{2}\, \frac{e^{-2u}}{(e^{-2u} + \overline{m}^2)}.
\eeq
As a consequence, both $\Phi_k(u)$ and $g_k(u)$ are real and have no explicit dependence on $k$ (or $x$). Said that, our aim now is to interpret (\ref{eq2.6}) as the action of a gravitational theory. To this end, let us first notice that according to (\ref{eq2.7}),
\beq
\label{eq2.10}
e^{-u}\, \partial_u\, \Phi(u) = \sqrt{g_{uu}}\, e^{-u}=\sqrt{g}, 
\eeq
where $g={\rm det}\, g_{ab}$ with $g_{ab}={\rm diag}\, \lbrace\, g_{uu},\, e^{-2u}\, \rbrace$. With this, one may formally write, 
%$\mathit{S}_{{\rm eff}}[\Phi,\overline{\Phi}]$ as,
\beq
\label{eq2.11}
%\mathit{S}_{{\rm eff}}[\Phi,g] = -2 \int d^2\sigma\, \sqrt{g}\, e^{u}\, \overline{\Phi}(u),
\mathit{S}_{{\rm eff}}[\Phi,g] =\frac{1}{4} \int d^2\sigma\, \sqrt{g}\, \mathcal{R}^{(2)}\, \Phi(u),
%\mathit{S}_{{\rm eff}}[\Phi,g] = \int d^2\sigma\, \sqrt{g}\, \mathcal{R}^{(2)}\,  \Phi(u),
\eeq
with $d^{\, 2}\sigma = du\, dx$ and  $\mathcal{R}^{(2)}=-8$ corresponding to the scalar curvature of the metric tensor $g_{ab}$.  The explicit dependence of $\mathit{S}_{{\rm eff}}[\Phi,g]$ on both $g_{ab}$ and $\Phi(u)$ in (\ref{eq2.11}), is to suggest that, despite these quantities are related by $\sqrt{g_{uu}}=\partial_u\Phi(u)$, they could be treated as two independent dynamical variables under some circumstances. This point will be clarified in the next section.

Regarding (\ref{eq1.11x}) one notices that $\Phi(u) = \langle \Phi\, |\, K_{-}\, |\, \Phi \rangle$. This amounts to see $\Phi(u)$ as a condensate of bosonic scalar particles. As coherent states are \textit{the most classical} states of a quantum system, thus the expectation value of an operator in a basis of these states is expected to behave mostly as a classical variable. 
%as quantum dispersions of that operator are supressed by orders of $\hbar$. 
In addition, \cite{molina015} has argued that the field $\Phi(u)$ of the 2-dimensional effective theory (\ref{eq2.8}) may be understood as an information-theorethic quantity. Namely, it amounts to the entanglement entropy between the left and right moving modes $a_k$ and $a_{-k}$ needed to create the cMERA state (\ref{eq1.3}) at each length scale $u$.

\section{cMERA effective action and Two Dimensional String Theory}
\label{2dstringth}
The aim in the following is to show that $\mathit{S}_{{\rm eff}}[\Phi,g]$ may be interpreted as the \textit{dilaton} term  of the generalized nonlinear sigma model that describes the worldsheet action of strings moving on a curved background \cite{friedan,fradkin,callan}. Namely, we show that $\Phi(u)$ in (\ref{eq2.8}) corresponds to a known solution of the equations of motion of the background fields. The discussion above also indicates that the background fields of the string worldsheet action might be seen as the expectation values of field operators in the cMERA coherent state basis.

The nonlinear sigma model worldsheet action for a closed bosonic string is given by,
\beq
\label{eq3.1}
S_{\rm ws} = \frac{1}{4\pi\alpha'}\, \int_{\Sigma} d^{\, 2}\sigma\, \sqrt{g}\, \left[g^{ab}G_{\mu\nu}(X)\, \partial_a X^{\mu}\partial_b X^{\nu} \right] +\frac{1}{4\pi}\, \int_{\Sigma} d^{\, 2}\sigma\, \sqrt{g}\, \mathcal{R}^{(2)}\, \Phi(X), 
\eeq
where $\sigma$ and $g_{ab}$ are coordinates and the metric on the worldsheet respectively; $\mathcal{R}^{(2)}$ represents the corresponding scalar curvature; $X^{\mu}$ denote target space coordinates with $\mu = 0 \cdots D-1$ and $D$ the dimension of the target spacetime, $G_{\mu\nu}(X)$ is the target spacetime metric and $\Phi(X)$ the \textit{dilaton} field. For convenience, the antisymmetric $B$-axion field has been set to zero. As usual, $\alpha'$ is the inverse string tension. From the point of view of the 2D quantum field theory on the worldsheet, changing the \textit{background} fields $G_{\mu\nu}$ and $\Phi$ amounts to consider a different theory. From the full string theory perspective, this  merely means a different background (state) within the same theory. The consistency of the string requires the local scale invariance of the quantum field theory on the worldsheet. This imposes the vanishing of the trace of the 2D worldsheet energy-momentum tensor $T_{ab}$. To satisfy this constraint in non-critical dimensions, the metric on the worldsheet and the dilaton must be treated as independent quantum dynamical variables although in the classical theory $g_{ab}$ depends on $\Phi$. The condition $T^{a}_a=0$ is accomplished by the vanishing of the the non linear sigma model $\beta$-functions $\beta^{\, G}_{\mu \nu}$ and $\beta^{\, \Phi}$ that, at one loop in $\alpha'$ are,
\barray
\label{eq3.2}
\frac{\beta^{\, \Phi}}{\alpha'}&=&\frac{(D-26)}{6\, \alpha'} + \frac{1}{2}\left[4(\nabla_{\mu}\Phi)^2 -2\nabla^{2}\Phi - \mathcal{R} \right], \\ \nonumber
\beta^{\, G}_{\mu \nu}&=& \mathcal{R}_{\mu\nu} + 2\nabla_{\mu}\nabla_{\nu}\Phi,
\earray
where $\nabla_{\mu}$ corresponds to the spacetime covariant derivative and $\mathcal{R}$ to the scalar curvature of the target spacetime. The vanishing of  $\beta^{\, G}_{\mu \nu}$ and $\beta^{\, \Phi}$ leads to the effective equations of motion for the background fields $G_{\mu \nu}$ and $\Phi$. 

\subsection{Linear Dilaton Background}
\label{Lin_Dil}
A consistent background solution to the equations of motion (\ref{eq3.2}) for arbirtary $D$ consists in a flat target spacetime and a \textit{linear dilaton},
\beq
\label{eq3.3}
G_{\mu\nu}(X)=\eta_{\mu\nu},\quad \Phi(X)=V_{\mu}\, X^{\mu},\, \quad V_{\mu}V^{\mu}=\frac{(26-D)}{12\alpha'}.
\eeq
For $D < D_{\rm crit}=26$, the dilaton gradient is spacelike. 

Here, we will consider the case for $D=2$ and $\Phi(X)$ lying along $X^{1}$, i.e, $\Phi(X)=Q\, X^{1}$ and $Q^2 = 2/\alpha'$. Thus, the geometry seen by the propagating string is a two dimensional flat spacetime with a dilaton linearly varying along its direction $X^{1}$. This is tantamount to say that the strength of the string interactions varies as a function of the $X^{1}$ coordinate. Indeed, the dilaton field \textit{defines} a coupling constant,
\beq
\label{eq3.4}
g_{\rm eff}= e^{\, \Phi(X)}.
\eeq 
Then, in the linear dilaton background where $\Phi = QX^{1}$, in the $X^{1}\to \infty$ region of the target spacetime $g_{\rm eff}=e^{QX^{1}}$ diverges and string perturbation theory fails. 

\subsection{cMERA Linear Dilaton}
\label{cMERA_lindil}
With this, let us now analyze the variational cMERA solution to $\Phi$ given in (\ref{eq2.8}) when $\overline{m}=0$, i.e, when considering the free massles scalar theory. In this case,
\beq
\label{eq3.5}
\Phi(u) = Q u \quad {\rm with} \quad  Q=1/2,
\eeq
and $g_{uu} =  1/4$. Choosing the target spacetime coordinates\footnote{We are working on a Euclidean signature. The Minkowski case may be obtained by putting $x \to i\theta$.} as $X^{\mu}=(X^{0}, X^{1})=(x, u)$ and $G_{\mu\nu}(X)={\rm diag}\,\lbrace 1, g_{uu} \rbrace$, the cMERA effective action reads as,
\beq
\label{eq3.6}
\mathit{S}_{{\rm eff}} =\frac{1}{4} \int d^2\sigma\, \sqrt{g}\, \mathcal{R}^{(2)}\, Q X^{1}.
\eeq 
In words, when $\overline{m}=0$, it is suggested that the cMERA effective action describes a linear dilaton background with $Q=1/2$. It must be noted that the consistency condition $Q^2 = 2/\alpha'$ imposes that one has to work in units where $\alpha'  \equiv 8$. This amounts to define a fiducial string interaction strength $g^2_{0}\sim \alpha'$ which landmarks the regime  $g_{\rm eff} \ll g_{0}$ where perturbation theory is valid. In the two dimensional string theory provided by the cMERA action (\ref{eq3.6}), $g_{\rm eff}^{2}=e^{2QX^{1}} = e^{u}$ so perturbation theory is valid for small values of the $u$-coordinate. There $g_{\rm eff}^{2} \approx 1 \ll g_0$. From the entanglement renormalization point of view this translates into the following: cMERA states $|\Phi\rangle$ close to the UV point, i.e, those in which the entanglement at all lengths scales has been added, correspond to regions of the \textit{"dual"} target spacetime where string perturbation theory is valid. On the other hand, as one runs into the IR region of cMERA, the resulting states have been devoid of their entanglement at small length scales. This region where a significant amount of short range entanglement has been discarded corresponds to the strong coupling region of the cMERA linear dilaton background. Furthermore, the inverse string coupling limits the number of the left-right moving entangled modes at the scale $u$ to those with momentum $k \leq \Lambda\, g_{\rm eff}^{-2}$. 

Now, in order to analyze the $\overline{m}\neq 0$ case, let us first review on another non-trivial background of two dimensional string theory.

\subsection{Two dimensional Black Hole background}
\label{2dbh}
As a theory of quantum gravity, string theory is also able to describe settings involving strong gravitational fields like black holes. An example of a well known non trivial solution to the background field equations (\ref{eq3.2}) is that of the black hole in two spacetime dimensions \cite{witten91}. In this solution, the spacetime manifold can be actually seen as parametrizing the coset group ${\rm SL}(2,\mathbb{R})/U(1)$. In Euclidean signature, with $X^{1} \geq 0$ and making $X^{0}$ periodic, the spacetime geometry seen by the string has the shape of a \textit{cigar}. The non-trivial fields in spacetime are the metric and the dilaton given by,
\barray
\label{eq3.7}
%G_{00}&=& 1,\, \quad G_{11}= \varrho^2\tanh^{2}(2 Q\, X^{1} + \log M), \\ \nonumber
G_{11}^{bh}(X) & = & \frac{1}{4}\tanh^{2}(2 Q\, X^{1} + \log M), \\ \nonumber
\Phi_{bh}(X^1) & = & -\frac{1}{2}\log 2 M -\frac{1}{2}\log\, \cosh\, (2 Q\, X^{1} + \log M),
\earray
with $G_{00}^{bh}(X)=1$ and $M$ being a mass constant. As $M \to 0$, the background (\ref{eq3.7}) approaches (\ref{eq3.3}). The linear dilaton is also recovered when $X^{1} \to 0$ and then the manifold resembles a cylinder instead of a cigar. Apart from being of great interest as a black hole, this solution arises in many other contexts in string theory, for example as the near-horizon limit of NS5-branes \cite{seiberg}. There, authors have also considered these linear dilaton backgrounds from the holographic point of view.

\subsection{cMERA 2D Black Hole}
\label{cMERA_bh}
Now we turn to analyze the variational solution (\ref{eq2.8}) for non zero mass. Regarding $\overline{m}\ll 1$, one can write,
\beq
\label{eq3.8}
\Phi(u) = Qu -\frac{1}{4}\log\left(1+\overline{m}^{\, 2}\, e^{2 u} \right) + \mathcal{O}(\overline{m}^{\, 2}), \quad Q=1/2,
\eeq 
and
\beq
\label{eq3.9}
g_{uu} = g(u)^2=\left[ \partial\, \Phi(u)\right] ^2 = \frac{1}{4}\left(1- \frac{\overline{m}^{\, 2}}{(\overline{m}^{\, 2} + e^{-2u})}\right)^2. 
\eeq 
Again, we choose the target spacetime coordinates as $X^{\mu}=(X^{0}, X^{1})=(x, u)$ and $G_{\mu\nu}(X)={\rm diag}\,\lbrace 1, g_{uu} \rbrace$. Two interesting limits may be identified. First of all we consider the $X^1 \to 0$ one. In this case,
\beq
\label{eq3.10}
\Phi(X^1) \approx Q X^1, \quad G_{11}(X) \approx \frac{1}{4}\, (1 - 2\overline{m}^{\, 2} e^{2 X^1}) \approx \frac{1}{4},
\eeq
thus recovering the cMERA linear dilaton background where $g_{\rm eff}^{2} = e^{2 Q X^1} \sim 1$. In order to justify the second limit, notice that the scalar curvature of $g_{ab}$,
\beq
\label{eq3.11}
\mathcal{R}^{(2)}=-8 + 8\overline{m}^{\, 4} e^{4X^1},
\eeq
remains constant along the $X^1$ coordinate before it exponentially vanishes when reaching $X^1_{*}\sim -\log \overline{m}$. We interpret this as a breakdown of the linear dilaton behaviour at $X^1_{*}$. Namely, at this value of the $X^1$-coordinate,
\beq
\label{eq3.12}
\Phi(X^1_{*}) \approx -\frac{1}{2}\log \overline{m},
\eeq
while $G_{11}$ in (\ref{eq3.10}) changes its sign, which might be interpreted by the presence of an horizon. In this limit, the effective string coupling $g_{\rm eff}^{2} = e^{2 Q X^1_{*}} \sim 1/\overline{m}\gg g_0$, which is far from the perturbative regime. One notices that the behaviour of these cMERA \textit{background fields} can be fairly accounted in terms of a two dimensional black hole solution (\ref{eq3.7}) by taking $Q=1/2$ and $M\equiv \overline{m}$, i.e,
\barray
\label{eq3.13}
G_{11}^{bh}(X) & = & \frac{1}{4}\tanh^{2}(X^{1} + \log \overline{m}), \\ \nonumber
\Phi_{bh}(X^1) &=& -\frac{1}{2}\log 2 \overline{m} -\frac{1}{2}\log\, \cosh\, (X^{1} + \log \overline{m}) \\ \nonumber
&=& QX^1 - \frac{1}{2}\log(1+ \overline{m}^{\, 2}e^{2X^1}).
\earray
With this identification one easily recovers the asymptotic value of the cMERA \textit{fields} for $X^1 \to 0$,
\beq
\label{eq3.14}
\Phi_{bh}(X^1) \approx Q X^1, \quad G_{11}^{bh}(X) \approx \frac{1}{4}\, (1 - 2\overline{m}^{\, 2} e^{2 X^1}) \approx \frac{1}{4},
\eeq
while $\Phi_{bh}(X^1_{*}) \approx -\frac{1}{2}\log \overline{m}$. 

 In view of these results, one might think in reverse and come to conclude that the two dimensional string theory linear dilaton, actually provides the solution for the $g(u)$ of the massive free boson as one considers the cMERA-like relation,
\beq
\label{eq3.15}
g(u)=\left[ \partial_{X^1}\, \Phi_{bh}(X^1)\right]_{X^1=\, u} =  -\frac{1}{2}\tanh(u + \log \overline{m}),
\eeq
which asymptotes to (\ref{eq2.9}) when $u \ll -\log \overline{m}$. 

In regard to this, if it is assumed that a linear dilaton background may describe the massles scalar cMERA, one could ask which value of $Q$ would make the correspondence to work. To this end we note on how two point functions are computed by cMERA. Given a scaling operator $\mathcal{O}$ of the theory, cMERA states that $\langle \Psi(0)\, |\, \mathcal{O}\, |\, \Psi(0)\rangle = \langle \Psi(u)\, |\, \mathcal{O}(u)\, |\, \Psi(u)\rangle$ with  $\mathcal{O}(u) =  U^{-1}(0,u)\, \mathcal{O}\, U(0,u)$ and $U(0,u)=\exp\lbrace -i\int_{u}^{0} d\hat{u}\, (\mathcal{K}(\hat{u}) + \mathcal{D})\rbrace$. In the scalar free theory, the simplest scaling operators are $\phi(x) \to e^{-\Delta_{\phi}}\, \phi(xe^{-u})$ and its conjugate momentum $\theta(x)\to  e^{-\Delta_{\theta}}\, \theta(xe^{-u})$ with $\Delta_{\phi}=0$ and $\Delta_{\theta}=1$ respectively. Noting that \cite{taka1},
\beq
\label{eq3.17}
U^{-1}\, \theta(x)\, U = e^{ -\Phi(u)-u/2}\, \theta(xe^{-u}),
\eeq
then,
\barray
\label{eq3.18}
 \langle \Psi(0)\, |\, \theta(x)\, \theta(x')\, |\, \Psi(0)\rangle  =   \\ \nonumber
e^{ -2\Phi(u) - u }\, \langle \Psi(u)\, |\, \theta(xe^{-u})\, \theta(x'e^{-u})\, |\, \Psi(u)\rangle.
\earray
This expression is evaluated at the length scale $u_0 = \log|\, x-x'\, |$ where the coarse grained distance between the UV locations of the operators has shrinked to one in units of the cutoff distance and  $\langle \Psi(u_0)\, |\, \theta(xe^{-u_0})\, \theta(x'e^{-u_0})\, |\, \Psi(u_0)\rangle \sim$ constant. With this,
\barray
\label{eq3.19}
 \langle \, \theta(x)\, \theta(x')\,\rangle  \propto  e^{-2\Phi(u_0) - u_0} =  e^{ -(2Q + 1 )\log|\, x-x'\, |}\\ \nonumber 
 = |\, x-x'\, |^{-(2Q + 1)}.
\earray
As $\theta(x)$ is scaling operator with $\Delta_{\theta}=1$, its two point function reads as $\langle \, \theta(x)\, \theta(x')\,\rangle \propto |\, x-x'\, |^{-2\Delta_{\theta}}$. This allows one to write the on shell-like relation,
\beq
\label{eq3.20}
2\Delta_{\theta}= 2 Q + 1,
\eeq
which inmediatly fixes $Q$ to its expected value of $1/2$.

Finally, let us briefly discuss on the results presented in this Letter. The emergence of a subcritical string theory in two spacetime dimensions from the entanglement renormalization of a one dimensional free boson seems puzzling at least. One might also wonder about how general this scheme can be. First we note that a non-critical string theory in two spacetime dimensions is a $c=1$ Liouville field theory whose Liouville field $\varphi_{L} \sim \Phi \sim X^{1}$ \cite{polyakov81}. It is also known that, through a B\"acklund transformation it is possible to map the Liouville field $\varphi_L$ onto a free field theory \cite{braaten82}. On the other hand, as it has been pointed out above, the cMERA flow for the free boson amounts to a coherent state \textit{evolution} with a kernel given by equations (\ref{eq2.2}) and (\ref{eq2.3}). Remarkably, in \cite{dariano85}, authors showed how these kind of coherent state evolutions can be thought as a succesion of infinitesimal local B\"acklund transformations. These facts together suggest that the cMERA flow of a free scalar field could effectively implement a  B\"acklund transformation and yield a dual description in terms of a subcritical string theory. It would be desirable to explicitly check this proposal in subsequent works. This could also provide new insights on the nature of true degrees of freedom in two dimensional string theory.

\section{Conclusions}
We have shown how the cMERA representation of different ground states of the free scalar boson correspond to non-trivial backgrounds of two dimensional string theory. This brings up the question if another non-trivial backgrounds may consistently be adscribed to different kind of states of the theory apart from the ground states. In addition, we have provided some insights on how the background fiels  $G_{\mu\nu}$ and  $\Phi$ arise from the structure of the entanglement between the left and right moving modes  that builds up the cMERA renormalization flow.  The entanglement entropy in the linear dilaton background of 2D-string theory was computed some time ago using its dual $c=1$ matrix model \cite{das1,das2} and more recently in \cite{hartnoll}. It would be worth to investigate to which amount this kind of entanglement can be related with the entanglement obtained there.

\section*{Acknowledgements}
This work has been supported by  Ministerio de Econom\'ia y Competitividad of Spain Project No. FIS2012-30625. Author also thanks the partial support of the European Science Fundation HoloGrav Network.

\end{document}